\documentclass[reprint,
nofootinbib,
amssymb,
aps,
]{revtex4-1}

\usepackage[intlimits]{amsmath}

\usepackage{graphicx}
\usepackage{dcolumn}
\usepackage{bm}
\usepackage[english]{babel}

\usepackage{tikz}
\usepackage{tikz-3dplot}
\usepackage{pgfplots}
\usepackage{wrapfig}
\usetikzlibrary{patterns}
\usetikzlibrary{decorations.markings}
\usepackage{verbatim}

\usepackage[colorlinks=true,linkcolor=blue, urlcolor= magenta, citecolor = magenta]{hyperref}

\begin{document}

\title{Mean lifetime of a false vacuum in terms of the Krylov-Fock non-escape probability}

\author{Michael~Maziashvili}
\email{maziashvili@iliauni.edu.ge}
\affiliation{School of Natural Sciences and Medicine, Ilia State University,\\ 3/5 Cholokashvili Ave., Tbilisi 0162, Georgia}


\begin{abstract}
The Krylov-Fock expression of non-decay (or survival) probability, which allows to evaluate the deviations from the exponential decay law (nowadays well established experimentally), is more informative as it readily provides the distribution function for the lifetime as a random quantity. Guided by the well established formalism for describing nuclear alpha decay, we use this distribution function to figure out the mean value of lifetime and its fluctuation rate. This theoretical framework is of considerable interest inasmuch as it allows an experimental verification. Next, we apply the Krylov-Fock approach to the decay of  a metastable state at a finite temperature in the framework of thermo-field dynamics. In contrast to the existing formalism, this approach shows the interference effect between the tunnelings from different metastable states as well as between the tunneling and the barrier hopping. This effect looks quite natural in the framework of consistent quantum mechanical description as a manifestation of the "double-slit experiment". In the end we discuss the field theory applications of the results obtained.   
 

\end{abstract}

\pacs{Valid PACS appear here}
\maketitle
\tableofcontents

\section{Introductory remarks}

The phenomenon of false vacuum decay plays an important role in evolution of the universe from its early beginnings to the present state \cite{Peacock:1999ye, Sher:1988mj}. Much of our understanding of tunneling, which is one of the basic ways for the false vacuum decay, comes from the one-dimensional quantum mechanics. Namely, for handling the false vacuum problem in quantum field theory, one usually reduces the problem to the one-dimensional case by using the JWKB approximation and the notion of a most-probable escape path. Most of the calculations in the case of field theory is reduced to the evaluation of JWKB tunneling probabilities. There is, however, a number of essential features concerning the calculations of mean-lifetime that lack a desirable transparency even in quantum mechanics. Fist of all, the lifetime of a metastable state is a random quantity and consequently for estimating the mean-lifetime one needs to know the distribution function for this quantity. Secondly, using this distribution function, one has to work out the fluctuation rate of the lifetime to have a "complete" description of the phenomenon. Therefore, we must step back and explain those features first in quantum mechanics. Our fundamental tool will be a none-escape (or survival) probability, $\omega(t)$, introduced by Krylov and Fock \cite{1947ZhETF..17..193K, Fock:2004mm}, which enables one to describe the whole dynamics of the unstable state decay. Presently, we know that the temporal development of the decay of a meta-stable-state manifests the presence of three regimes: initially decay is slower than exponential; then comes the exponential decay, which for the long-times is followed by the inverse power law \cite{1957SPhD....2..340K, 1958JETP....6.1053K, 1960SPhD....5..515K, Fonda:1978dk}. The existence of these three regimes appears to be an universal feature of the decay process.\footnote{The only exception of which we are aware is the escape of massive scalar particles from the brane. In this particular case the decay strictly follows an exponential law \cite{Maziashvili:2005cd}.} The attempt to use $\omega(t)$ for a better estimate of the mean lifetime of a false-vacuum was made in a few years ago in \cite{Andreassen:2016cff, Andreassen:2016cvx}. Their approach does not address the questions posed above but rather is aimed to extract $\Gamma$ factor that governs the decay at intermediate time-scales. It is also worth mentioning, that the first (perhaps not very successful) attempt to generalize the quantum-mechanical results of non-exponential decay to the field theory was made in \cite{Krauss:2007rx}.

We begin by introducing the distribution function for the lifetime. It is based on the Krylov-Fock non-escape probability. After discussing the mean-lifetime and its fluctuation rate, we address the decay of an unstable system at a finite temperature. For this purpose, we use the formalism of thermo-field dynamics, which allows the distribution function for the lifetime to be introduced in the same manner. In the end, we discuss the applications to the field theory and summarize our results. Many results of the discussion can be checked experimentally.

\section{Krylov-Fock survival probability and mean-lifetime}

\begin{figure}[h]
	\centering
	
	\begin{tikzpicture}
	
	\draw[->] (-0.3,0) --  (7, 0) node[anchor=north east]{$x$};
	\draw[->] (0,-0.3) --  (0, 4.5) node[anchor=north east]{$\Phi(x)$};
	
	
	
	\draw [brown, thick] plot [smooth] coordinates {(0.2,4.4) (0.5,0.9) (1.4,0.9) (2,3.5) (3,3.5)   (4,1.2)  (5,0.4) (6.8,0.23)}; 
	
	\draw [blue, dashed, thick] (2.44,3.75) -- (2.44,0) node[black, below]{$x_{top}$} ; 
	
	\draw [blue, dashed, thick] (0.88,0.59) -- (0.88,0) node[black, below]{$x_{min}$}; 
	
	\draw[blue,  dashed, thick] (0,0.9) node[black, align=left] {$\mathcal{E}_0~~~~$} -- (1.46,0.9) ;
	
	\draw[blue, dashed, thick] (0,1.34) node[black, align=left] {$\mathcal{E}_1~~~~$} -- (1.55,1.34) ;
	
	\draw[blue, dashed, thick] (0,1.9) node[black, align=left] {$\mathcal{E}_2~~~~$} -- (1.66,1.9) ; 
	
	\draw[blue, dashed, thick] (0,2.6) node[black, align=left] {$\mathcal{E}_3~~~~$} -- (1.76,2.6) ;

	
	
	
	\end{tikzpicture}

		\caption{Schematic drawing of the potential.} \label{fig1}
\end{figure}
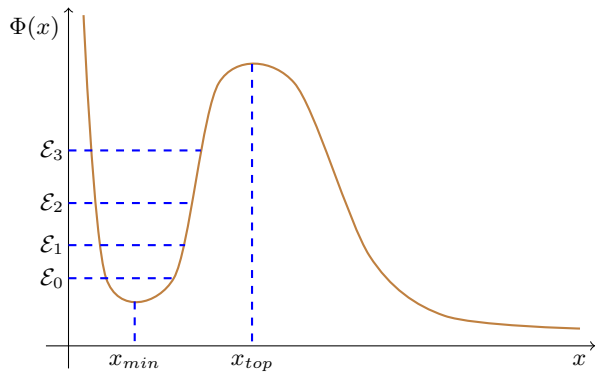

The temporal development of the decay of a metastable state is conveniently described in terms of a non-escape/survival probability

\begin{eqnarray}\label{krilovfoki}
w(t) = |\langle \psi_{in}|\psi(t)\rangle|^2 ~, 
\end{eqnarray} where $\psi_{in}(x)$ is the wave function describing the particle confined at $t=0$ to the potential well - whose motion is inhibited by the potential barrier, see Fig.\ref{fig1} - and $\psi(t)$ is a solution of the Schr\"odinger equation with this initial state.

\noindent  On the basis of this formalism, put forward by Krylov and Fock \cite{1947ZhETF..17..193K, Fock:2004mm}, one can gain a qualitative understanding of the whole evolution of a metastable state. Let us consider a schematic potential shown in Fig.\ref{fig1}. It tends to infinity as $x$ tends to $0^+$, has a local maximum at $x_{top}$ and then decays monotonically for $x > x_{top}$ - tending asymptotically to $0$. For such potential, the Hamiltonian has a continuous energy spectrum, $\psi_{\mathcal{E}}(x), 0 < \mathcal{E} <\infty$, \begin{eqnarray}
\int_{-\infty}^\infty \mathrm{d}x \, \psi_{\mathcal{E}}^*(x) \psi_{\mathcal{E}'}(x) \,=\, \delta\big(\mathcal{E} \,-\, \mathcal{E}'\big) ~, \nonumber 
\end{eqnarray} which allows to express the amplitude $\langle \psi_{in}|\psi(t)\rangle$ as 

\begin{eqnarray}\label{amplitude}&& \langle \psi_{in} |\psi(t) \rangle \,=\,   \int_0^\infty \mathrm{d}\mathcal{E} \, \mathrm{e}^{-it\mathcal{E}} |\langle \psi_{in}|\psi_{\mathcal{E}} \rangle |^2~. \end{eqnarray}  

Whenever the potential is bounded from below, the decay for $t\to\infty$ follows not exponential but rather an inverse power law \cite{1957SPhD....2..340K, 1958JETP....6.1053K, 1960SPhD....5..515K} (has been confirmed experimentally \cite{2006PhRvL..96p3601R, Kelkar:2004zz}). On the other hand, the decay is initially slower than that predicted by the exponential law \cite{Fonda:1978dk} (observed experimentally \cite{1997Natur.387..575W, 2001PhRvL..87d0402F}). It can be readily seen by estimating $\omega(t)$ for small values of $t$: $\omega(t) = 1 - t^2 \delta E_{\psi_0}^2 + O(t^4)$, where $\delta E_{\psi_0}$ is the variance (or fluctuation) of energy. In the case of exponential decay, $\omega(t) = \exp(-t/\tau)$, one would have: $\omega(t) = 1 - t/\tau + O(t^2)$. Thus, at short and long times there are deviations from the exponential decay law.

The question we want to address now is as follows. What does the function \eqref{krilovfoki} signify? The physical significance of $\omega(t_0)$ is that it defines the probability of finding a state $\psi(t)$ initially confined to the region of a potential-well in that same region after a time $t_0$. The lifetime is a random variable and to characterize it we need the distribution function. Using the terminology well established in probability theory \cite{1962PhT....15j..62G, 1971aitp.book.....F}, $\omega(t)$ is the distribution function of lifetime of the metastable state $\psi_{in}$. That is, $\omega(t_0)$ stands for probability that the metastable state will survive for the time interval $0\leq t\leq t_0$, or equivalently, that the lifetime, $t$, will be greater than $t_0$. In other words, the probability of finding lifetime in the time interval $t_1 < t < t_2$ is given by

\begin{eqnarray}
	W\{t_1 < t < t_2\} \,=\, - \int_{t_1}^{t_2}\mathrm{d} \omega \,=\, \omega(t_1) \,-\, \omega(t_2) ~. \nonumber 
\end{eqnarray} It implies that $-\dot{\omega}$ is a probability density of the lifetime. That is, $-\mathrm{d}\omega$ gives the probability that the system will not decay from $t$ to $t+\mathrm{d}t$. This definition meets the requirement that the total probability is unity

\begin{eqnarray}
- \, \int_0^\infty \mathrm{d} \omega \,=\, \omega(0) \,-\, \omega(\infty) \,=\, 1 ~. \nonumber 
\end{eqnarray} Thus, the mean lifetime can be estimated as 

\begin{eqnarray}\label{Lebensdauer}
\tau \,=\, -\int_0^\infty \, t \mathrm{d}\omega(t) \,=\, \int_0^\infty \mathrm{d}t \,\omega(t) \,-\, \lim_{t\to \infty} t\omega(t)  ~. 
\end{eqnarray}

Let us note that the definition of mean-lifetime by the Eq.\eqref{Lebensdauer} is not new, for instance one can find similar discussion in \cite{Shirokov}, but is not a common knowledge. The mean-lifetime is often defined in the literature as (see \cite{Valentin, Krane:1987ky, Nicolaides}, possibly one can add many other references)

\begin{eqnarray}\label{AlterAusdruck}
	\widetilde{\tau} \,=\,  \left.\int_0^\infty \mathrm{d}t \,t\omega(t)\right/\int_0^\infty \mathrm{d}t \,\omega(t) ~. 
\end{eqnarray} 

 We adopt the Eq.\eqref{Lebensdauer} as the correct one but strictly speaking only the experiment can decide the question. Both expressions give precisely the same results for the exponential decay 
 
 \begin{eqnarray}\label{sameresults}
 	\omega(t) = \exp(-t\Gamma) ~, ~~\Rightarrow~~ \tau = \widetilde{\tau} = \frac{1}{\Gamma} ~. 
 \end{eqnarray} However, the decay does not precisely follow the exponential law and one may hope that the corrections due to deviation from the exponential law will allow one to distinguish between these two expressions. For this purpose, in the next section we shall consider a typical quantum-mechanical example that can serve as a guide to experiments that might be sensitive to such corrections.

\section{Cauchy-Lorentz distribution}
\label{Cauchy-Lorentz}

Let us consider one of the typical examples of energy distribution for the metastable state (some other examples can be found in \cite{Douvropoulos}) in order to estimate the mean-lifetimes with respect to Eq.\eqref{Lebensdauer} and Eq.\eqref{AlterAusdruck}. For narrow resonances, $\Gamma/\mathcal{E}_0\ll 1$, the integrand in Eq.\eqref{amplitude} is usually approximated by the Cauchy-Lorentz (Breit-Wigner) distribution \cite{1959HDP....41....1B}

\begin{eqnarray}\label{Breit-Wigner}
	|\langle\psi_0|\psi_{\mathcal{E}} \rangle |^2 \,=\, \frac{N}{2\pi} \,  \frac{1}{\left(\mathcal{E} - \mathcal{E}_0\right)^2 + \Gamma^2/4} ~, \nonumber 
\end{eqnarray} where $N$ stands for the normalization factor to ensure

\begin{eqnarray}
\langle\psi_0|\psi_0\rangle \,=\,  \int_0^\infty\mathrm{d}\mathcal{E} \, |\langle\psi_0|\psi_{\mathcal{E}} \rangle |^2 \,=\, 1~ \Rightarrow~  \nonumber \\ N = \frac{2\pi\Gamma}{1-\pi +2\arctan(2\mathcal{E}_0/\Gamma)} ~. \nonumber
\end{eqnarray} 

For the mean lifetimes one obtains (see Appendix)

\begin{eqnarray}
&&\tau =   \frac{N^2}{4\pi^2\Gamma^3} + \frac{N^2(\pi-1)}{\pi^2\Gamma^3}\left(\frac{\pi}{2}-\arctan(\xi_0) - \frac{\xi_0}{1+\xi_0^2}\right)   ~, \nonumber \\&&
\widetilde{\tau}\tau = \frac{N^2}{4\pi^2\Gamma^4} - \frac{4N^2}{\pi^2\Gamma^4}\left(\frac{\pi}{8} -\frac{\xi_0(3+\xi_0^2)}{4(1+\xi_0^2)^2} - \frac{\arctan(\xi_0)}{4} + \right. \nonumber \\&& \left. \left(\frac{\pi}{2} - \arctan(\xi_0)\right)\frac{2\xi_0}{1+\xi_0^2} \right) ~, ~~\text{where} ~~  \xi_0 \equiv 2\mathcal{E}_0/\Gamma \nonumber
\end{eqnarray}

In order the measurements to reveal the difference between $\tau$ and $\widetilde{\tau}$ it is necessary the width $\Gamma$ not to be very small compared to $\mathcal{E}_0$ - since otherwise the decay will basically follow the exponential law resulting in the equality $\tau = \widetilde{\tau}$.

\section{Fluctuations of the lifetime}
\label{fluctuations}

Along the mean lifetime, the probability density allows one to define its fluctuation rate as well

\begin{eqnarray}\label{Schwankung}
\delta \tau \,=\, \left(-\int_0^\infty \, t^2 \mathrm{d}\omega(t) -\tau^2\right)^{1/2} \,=\, \nonumber \\ \left( 2\int_0^\infty \, t\omega(t) \mathrm{d}t \,-\, \lim_{t\to \infty} t^2\omega(t) \,-\, \tau^2  \right)^{1/2} ~. 
\end{eqnarray} In view of Eq.\eqref{AlterAusdruck}, the distribution function for the lifetime is understood to be

\begin{eqnarray}
\omega(t)\left/\int_0^\infty \mathrm{d}t \,\omega(t)\right. ~, \nonumber
\end{eqnarray} and for the fluctuations one would have

\begin{eqnarray}\label{AlteSchwankung}
\delta\widetilde{\tau} =  \left( \left.\int_0^\infty \mathrm{d}t \,t^2\omega(t)\right/\int_0^\infty \mathrm{d}t \,\omega(t) \,-\, \widetilde{\tau}\,^2 \right)^{1/2}  ~. 
\end{eqnarray} Both Eq.\eqref{Schwankung} and Eq.\eqref{AlteSchwankung} give the same results for the exponential decay

\begin{eqnarray}\label{expfluct}
	\omega(t) = \exp(-\Gamma t) ~,~~\Rightarrow~~ \delta\tau = \delta\widetilde{\tau} = \frac{1}{\Gamma}~. 
\end{eqnarray} We see that $\delta\tau(\delta\widetilde{\tau})$ turns out to be equal to $\tau(\widetilde{\tau})$, see \eqref{sameresults}. Of course, it may seem somewhat embarrassing in that when the fluctuations around the expectation value cannot be considered as negligible compared with the expectation value, then the latter is not very informative. To clarify this point, it maybe helpful to draw a simple parallel from quantum mechanics. Recall the ground state wave-function of a harmonic oscillator

\begin{eqnarray}
H =\frac{p^2}{2m} +\frac{m\omega^2(x-x_0)^2}{2} ~, \nonumber \\  \psi = \left(\frac{m\omega}{\pi}\right)^{1/4}\mathrm{e}^{-m\omega(x-x_0)^2/2} ~. \nonumber
\end{eqnarray} The average value of the position is clearly $\langle x\rangle = x_0$ while its fluctuation rate is $\delta x = 1/\sqrt{2m\omega}$. One can always arrange the parameters $x_0, m, \omega$ in such a way as to have $\langle x\rangle = \delta x$. Having ensemble of such oscillators, the measurement of the position would not give a sharp value. Similarly, if one considers the result \eqref{expfluct} to be fairly reliable, the measurement of the mean-lifetime for ensemble of radioactive nucleus would give quite deferent results spread over the region: $0 \lesssim  \tau \lesssim 2/\Gamma$. That means that the mean-lifetime is quite uncertain. Not to go astray, from this point on we proceed by recalling the standard approach to the alpha decay.

In order to compare with experiments, one is instead considering a huge number of identical metastable systems, $n_0$, and looks for the probability that $n$ particles out of $n_0$ will decay during the time $t$ \cite{Shirokov}

\begin{eqnarray}
W_n^{n_0}(t) = \frac{n_0!}{n!(n_0-n)!} \, \omega^{n_0-n}(t)\big(1-\omega(t)\big)^n ~.\nonumber
\end{eqnarray} Of course 

\begin{eqnarray}
\sum_{n=0}^{n_0} W_n^{n_0}(t) = 1 ~, \nonumber
\end{eqnarray} and one can use this distribution for estimating the average number of decaying particles and its fluctuation rate 

\begin{eqnarray}
\langle n \rangle(t) = \sum_{n=0}^{n_0} W_n^{n_0}(t)n ~, ~(\delta n)^2(t) = \langle n^2 \rangle(t) - \langle n \rangle^2(t) ~. \nonumber
\end{eqnarray} In the case of exponential decay, one usually assumes $n_0 \gg n$ and approximates $W_n^{n_0}(t)$ by the Poisson distribution. As a result, the average number of decay and its fluctuation rate appear to be 

\begin{eqnarray}\label{ShirYud}
	\langle n \rangle(t) \approx n_0\Gamma t ~, ~~ \delta  n(t) \approx \sqrt{n_0\Gamma t} ~. 
\end{eqnarray} If we had an ensemble of false vacua, say in the context of multi-verse, then this result might be of some use. To see how large the fluctuations in Eq.\eqref{ShirYud} may be, let us assume that we have such an ensemble. It is known that if the standard model of particle physics is valid all the way up to the Planck energy scale, then the lifetime of the electroweak vacuum is expected to be about $10^{655}$ times larger than the age of the universe $t_U$ \cite{Branchina:2014rva}. Putting in Eq.\eqref{ShirYud} $t=t_U$, then one would obtain that $\langle n \rangle(t)/\delta  n(t) \gg 1 \Rightarrow n_0\gg 10^{655}$. In fact, we are again facing the problem of large fluctuations. 

The problem of large fluctuations we encountered can be described in terms of time as well. To identify the mean value of $n$ in Eq.\eqref{ShirYud} when $1/\Gamma$ is large enough, one clearly needs to spend much time ($t>1/n_0\Gamma$). And in the case of above considered oscillator too, one has to make a huge number of measurements to identify the mean value of $x$ as the oscillator does not spend most time in the narrow region around the mean value.

Thus (for a false vacuum in the universe) we clearly face the problem of large $\delta\tau$ that remarkably reduces predictive power of $\tau$.

\section{False vacuum decay at a finite temperature }

Once one knows the zero-temperature description of the dynamics of a metastable state \cite{Fonda:1978dk}, an obvious
question arises - how to describe the same problem at a finite temperature? The main questions in dealing with this problem, however, are how do we choose initial state and how do we define the dynamical equation. Apart from the tunneling, the trapped particle experiencing thermal fluctuations can also hop over the barrier. At a certain temperature scale, the barrier hopping becomes dominant over the barrier-tunneling and, thus, it becomes reasonable to describe the process classically with small quantum fluctuations. The probabilistic description of the dynamics of particle hopping over the barrier, which is affected by the thermal and quantum fluctuations was suggested by Kramers \cite{Kramers:1940zz}. In his model, the distribution function, which obeys the Fokker-Planck equation, describes a large number of Brownian particles with no mutual interference - being initially in thermal equilibrium within the potential well. An alternate description is based on the use of Wigner function \cite{Calzetta:2006rg}. The equation of motion for the Wigner function looks like the Kramers (Fokker-Planck) equation with quantum corrections. Unfortunately, in course of time the Wigner function may become negative for some values of phase-space coordinates, even if it is initially positive-definite \cite{Pawula:1967zz, 1987ZPhyB..66..257R}. For this reason, in general, it cannot be regarded as the distribution function. As these approaches are of little use for our discussion, we will not discuss them into details. Instead we will approach this problem from the standpoint of thermo-field dynamics \cite{Takahashi:1996zn, Arimitsu:1985xn, Arimitsu:1985xm, Arimitsu:1985ez}.

\section{A glance at the thermo-field dynamics}

By now the formalism of thermo-field dynamics \cite{Takahashi:1996zn, Arimitsu:1985xn, Arimitsu:1985xm, Arimitsu:1985ez} is well established as a natural framework
to analyze time-dependent processes at finite temperature. It maybe of considerable interest for the problem under consideration. The key point in this formalism is the introduction of thermal vacuum state giving the expectation values equivalent to the Boltzmann averaging. It is a method for describing mixed states as pure states in an enlarged Hilbert space. To treat our problem within this formalism, we have to enlarge our system by introducing one more coordinate $y$. Applying Boltzmann averaging, one usually states that if a system in equilibrium can be in one of $\psi_j$ states, then the expected value of the observable is 

\begin{eqnarray}\label{Boltzmann}
\langle A \rangle_\beta = \frac{\sum_j\langle \psi_j|\widehat{A}|\psi_j\rangle\exp\big(-\beta\mathcal{E}_j\big)}{\mathcal{Z}} ~, 
\end{eqnarray} where $\beta\equiv k_BT$ and

\begin{eqnarray}
\mathcal{Z}\equiv \sum_j\exp\big(-\beta\mathcal{E}_j\big) ~. \nonumber 
\end{eqnarray} To represent the average $\langle A \rangle$ as the expectation value of the operator $\widehat{A}$ for a pure state, one constructs "thermal vacuum" in doubled Hilbert space

\begin{eqnarray}\label{thermalvacuum}
\psi_\beta(x,y) = \frac{1}{\sqrt{\mathcal{Z}}} \sum_j\exp\left(-\frac{\beta\mathcal{E}_j}{2}\right)\psi_j(x)\psi_j(y) ~.  
\end{eqnarray} As the operator $\widehat{A}$ acts in the $x$ space, one finds

\begin{eqnarray}
\langle\psi_\beta|\widehat{A}|\psi_\beta\rangle =\frac{1}{\mathcal{Z}}\sum_{i,j}\exp\left(-\frac{\beta\mathcal{E}_i}{2}\right)\left(-\frac{\beta\mathcal{E}_j}{2}\right)\times \nonumber \\  \iint\mathrm{d}x\mathrm{d}y \, \psi^*_i(y)\psi^*_i(x)\psi_j(y)\widehat{A}\psi_j(x) =  \nonumber \\ \frac{1}{\mathcal{Z}}\sum_{i,j}\exp\left(-\frac{\beta\mathcal{E}_i}{2}\right)\left(-\frac{\beta\mathcal{E}_j}{2}\right)\times \nonumber \\  \int\mathrm{d}y \, \psi^*_i(y)\psi_j(y)\int\mathrm{d}x\, \psi^*_i(x)\widehat{A}\psi_j(x) = \langle A \rangle_\beta ~. \nonumber 
\end{eqnarray} The reason why this approach maybe particularly useful in dealing with the decaying system is that usually the temperature is assumed to be low enough so that the initial state can be viewed as a near thermal equilibrium and can therefore be represented as a specific superposition of the metastable states. Having this sort of initial state, one can then proceed in the same fashion as in the zero-temperature case. That is, one could merely study the finite temperature decay dynamics by solving the Schr\"odinger equation with this initial state.

Assuming the sum in Eq.\eqref{thermalvacuum} is taken over the metastable states, the solution of the Schr\"odinger equation can be represented as

\begin{eqnarray}\label{evolutsia}
\psi_\beta(t, x,y) = \frac{1}{\sqrt{\mathcal{Z}}} \sum_j\exp\left(-\frac{\beta\mathcal{E}_j}{2}\right)\psi_j(y)\mathrm{e}^{-it\widehat{H}}\psi_j(x) ~.  \nonumber 
\end{eqnarray} Hence, for the temperature dependent trajectory of the escaping particle one would have 

\begin{eqnarray}
\langle x\rangle (t) = \frac{1}{\mathcal{Z}} \sum_j\exp\left(-\beta\mathcal{E}_j\right)\int\mathrm{d}x\, x|\psi_j(t,x)|^2~, \nonumber
\end{eqnarray} and somewhat similar expression for

\begin{eqnarray}
\langle p\rangle (t) = \frac{-i}{\mathcal{Z}} \sum_j\exp\left(-\beta\mathcal{E}_j\right)\int\mathrm{d}x\, \psi^*_j(t,x) \partial_x\psi_j(t,x) ~. \nonumber 
\end{eqnarray} Basically, it is tantamount to solving the Heisenberg equations for $\widehat{x}(t), \widehat{p}(t)$ operators and then taking Boltzmann averaging $\text{Sp}\Big(\widehat{\rho} \, \widehat{x}(t)\Big)$ and $\text{Sp}\Big(\widehat{\rho} \, \widehat{p}(t)\Big)$, where $\widehat{\rho} = \exp\left(-\beta\widehat{H}\right)$.

\section{Semiclassical approach}
\label{semiclassical}

Obviously, we have to be guided by the existing approach to the problem, which is experimentally verified. For this purpose we shall mainly use the textbook \cite{1950nuph.conf.....F}.

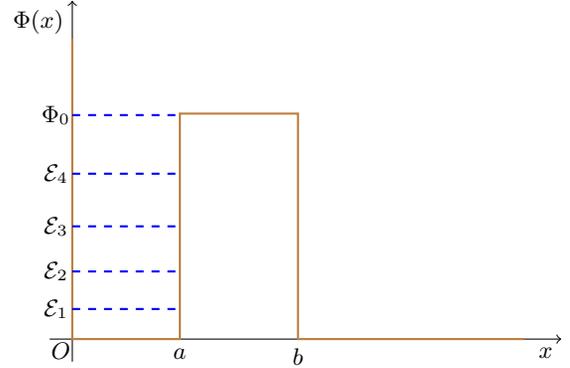
\begin{figure}[h]
	\centering

\begin{tikzpicture}

\draw[->] (-0.3,0) --   (6.5, 0) node[anchor=north east]{$x$};
\draw[->] (0,-0.3)  --  (0, 4.5) node[anchor=north east]{$\Phi(x)$};

\node[black] at (-0.15,-0.15) {$O$};




\draw [brown, thick]  {(0,4) -- (0,0)  -- (1.43,0) node[black, below]{$a$} -- (1.43,3) -- (3,3) -- (3,0) node[black, below]{$b$} -- (6,0) };

\draw[blue,  dashed, thick] (0,0.4) node[black, align=left] {$\mathcal{E}_1~~~~$} -- (1.43,0.4) ;

\draw[blue, dashed, thick] (0,0.9) node[black, align=left] {$\mathcal{E}_2~~~~$} -- (1.43,0.9) ;

\draw[blue, dashed, thick] (0,1.5) node[black, align=left] {$\mathcal{E}_3~~~~$} -- (1.43,1.5) ; 

\draw[blue, dashed, thick] (0,2.2) node[black, align=left] {$\mathcal{E}_4~~~~$} -- (1.43,2.2) ;    

\draw[blue, dashed, thick] (0,2.98) node[black, align=left] {$\Phi_0~~~~$} -- (1.43,2.98) ;




\end{tikzpicture}

	\caption{Rectangular potential.} \label{fig}
\end{figure}

In the case of a metastable system there is a finite number of virtual states - each of which is characterized with its lifetime $1/\Gamma_j$. If the potential barrier is large enough, one can find the virtual levels approximately by making the barrier infinitely wide. For simplicity, let us consider a rectangular version of the potential shown in Fig.\ref{fig1}, see Fig.\ref{fig}.

\noindent  Then, the metastable states can be defined approximately by the transcendental equation \cite{1965qume.book.....L}

\begin{eqnarray}\label{transtsedenturi}
a\sqrt{2m\mathcal{E}_j} = \pi j - \arcsin\sqrt{\frac{\mathcal{E}_j}{\Phi_0}} ~,  
\end{eqnarray} where the values of $\arcsin$ are taken between $0, \pi/2$. Making the assumption $\mathcal{E}_j/\Phi_0 \ll 1$, from Eq.\eqref{transtsedenturi} one finds

\begin{eqnarray}
\mathcal{E}_j = \frac{\pi^2j^2}{\Big(\sqrt{2m}a + 1/\sqrt{\Phi_0}\Big)^2} ~. \nonumber 
\end{eqnarray}  

Above the barrier, the energy spectrum of the particle is continuous - extending from $\Phi_0$ to infinity. This spectrum is given by the eigenvalues of the Hamiltonian. In general, the Hamiltonian with the potential shown in Fig.\ref{fig} possesses only continuous spectrum 

\begin{eqnarray}
H\psi(x, \mathcal{E})=\mathcal{E}\psi(x, \mathcal{E}) ~,~~ 0 <\mathcal{E} < \infty ~. \nonumber
\end{eqnarray} The virtual levels, $\mathcal{E}_j$, are distinguished by the fact that the function $|\psi(x, \mathcal{E})|$, is mainly concentrated within the potential well when $\mathcal{E}$ is close to $\mathcal{E}_j$. That is,

\begin{eqnarray}
\int_0^b\mathrm{d}x\, |\psi(x, \mathcal{E}_j)|^2 \gg  \int_x^{x+b}\mathrm{d}x\, |\psi(x, \mathcal{E}_j)|^2 ~, \nonumber 
\end{eqnarray} where $x$ is an arbitrary positive number obeying $x>b$.

The essential features of the metastable states, $\psi_j$, (not to be confused with $\psi(x, \mathcal{E}_j)$) are that they are normalized functions localized within the well by means of which the virtual energy levels are defined as: $\langle\psi_j|\widehat{H}|\psi_j\rangle = \mathcal{E}_j$. Let us assume that prior to decay the particle is in the state $\psi_1$. Now, the thing to compute is a decay probability per unit time: $\Gamma_1$. It is usually regarded as the product of tunneling probability, $W(\mathcal{E}_1)$, and the frequency by which a particle hits the barrier from inside \cite{1950nuph.conf.....F} 

\begin{eqnarray}\label{sikhshire}
\frac{v_1}{a} = \frac{\sqrt{2\mathcal{E}_1/m}}{a} ~. 
\end{eqnarray}

\noindent The process is qualitatively pictured as follows. The particle bounces around inside the well and, each time it hits the barrier, it has a certain probability of
tunneling through it. The Eq.\eqref{sikhshire} is nothing other than the probability that, per unit time, the particle will hit the starting point of tunneling. Thus, one arrives at the expression 

\begin{eqnarray}\label{gamaerti}
\Gamma_1 = \frac{\sqrt{2\mathcal{E}_1/m}}{a}\, W(\mathcal{E}_1) ~. 
\end{eqnarray} The next step is to recall that there is the probability

\begin{eqnarray}\label{energiisalbatobebi}
\rho(\mathcal{E})\mathrm{d}\mathcal{E} = |\langle\psi_1|\psi(\mathcal{E})\rangle|^2\mathrm{d}\mathcal{E} ~, 
\end{eqnarray} that the energy of particle will lie in the region $(\mathcal{E}, \mathcal{E}+\mathrm{d}\mathcal{E})$. The quantities \eqref{gamaerti} and \eqref{energiisalbatobebi} enable one to write

\begin{eqnarray}\label{ftmpep}
\Gamma_{tun}  = \frac{\sqrt{2/m}}{a} \int_0^{\Phi_0} \mathrm{d}\mathcal{E}\, \rho(\mathcal{E})\sqrt{\mathcal{E}}W(\mathcal{E}) ~. 
\end{eqnarray} 

The Eq.\eqref{energiisalbatobebi} indicates that the probability for $\mathcal{E}>\Phi_0$ is not zero. In such cases, the energy of particle is sufficient to surmount the barrier. That is, the over-barrier jumping takes place. How to estimate $\Gamma$ for this process? The answer can read off from the Eq.\eqref{gamaerti}. The lifetime of particle in the state $\psi_1$, which is the inverse of $\Gamma_1$, is the time $a/v_1$ amplified by the factor $1/W(\mathcal{E}_1)$. Thus, the factor $1/W(\mathcal{E}_1)$ has a clear physical meaning of the number of particle reflections from the barrier back to the well. Neglecting the over-barrier reflection, for the over-barrier jumping one obtains

\begin{eqnarray}\label{ftmph}
\Gamma_{ob}  = \frac{\sqrt{2/m}}{a} \int_{\Phi_0}^\infty \mathrm{d}\mathcal{E}\, \rho(\mathcal{E})\sqrt{\mathcal{E}} ~. 
\end{eqnarray} This equation allows a straightforward generalization to the finite temperature case by replacing

\begin{eqnarray}
\rho(\mathcal{E}) = \frac{\exp(-\beta (\mathcal{E} - \mathcal{E}_1))}{Z} ~. \nonumber
\end{eqnarray} With this replacement, the Eq.\eqref{ftmpep} becomes decay rate for the thermally assisted tunneling and can be approximated by the finite-temperature-most-probable-escape-path \cite{Weinberg:2012pjx, Noble:1981fh, Weinberg:2006pc, Brown:2007sd}. That means that instead of tunneling directly from the metastable state, the particle can tunnel from a thermally excited higher energy states. However, the evaluation of Eqs.(\ref{ftmpep}, \ref{ftmph}) is of little interest for our further discussion.

\section{Thermally mixed initial state}

Now let us return to the thermo-field dynamical description. To simplify ensuing discussion, we will again use the rectangular potential depicted in Fig.\ref{fig}. For further simplification, we will assume the existence of a single metastable state, which we denote by $\psi_0$. The energy spectrum above the barrier is denoted by $\psi_j$, where $j=1, 2, 3, \ldots$. The initial state, which is nearly normalized to unity, can be written as \eqref{thermalvacuum}

\begin{eqnarray}
\psi_\beta(x,y) = \frac{1}{\sqrt{\mathcal{Z}}} \sum_{j=0}^\infty\exp\left(-\frac{\beta\mathcal{E}_j}{2}\right)\psi_j(x)\psi_j(y) ~.  
\end{eqnarray} To clarify the point, we remind the reader that the initial state should be localized within the well: $0<x \lesssim a$. That means that all $\psi_j(x)$'s should be localized within the well. At the same time, we require that the energy spectrum above the barrier starts from $\Phi_0$. For constructing such an initial state in a simple way, one might start with the spectrum of an infinite well,

\begin{eqnarray}
\psi_j(x) = A_j\sin\left(\frac{\pi(j+1)x}{a}\right) ~, ~~ \mathcal{E}_j = \frac{\pi^2(j+1)^2}{2ma^2} ~, \nonumber 
\end{eqnarray} and then for $j\geq 1$ one could slightly adjust the parameter $a \to c$ in such a way as to ensure 

\begin{eqnarray}
\mathcal{E}_1	= \Phi_0 ~. \nonumber 
\end{eqnarray} Next, one can replace this spectrum by

\begin{eqnarray}
\psi_0 =\begin{cases}
\sqrt{\frac{1}{a}}\exp\left(i\frac{\pi x}{a}\right) ~, ~~ \text{for} ~~ 0<x<a ~, \\  ~~~~0 ~, ~~ \text{for}~~ x< 0 ~\text{and} ~ x>a ~,
\end{cases} ~~~~~ \nonumber \\ \label{barierzeda}
\psi_{j\geq 1} =  \begin{cases}
\sqrt{\frac{1}{c}}\exp\left(i\frac{\pi(j+1)x}{c}\right) ~, ~~ \text{for} ~~ 0<x<c ~, \\  ~~~~0 ~, ~~ \text{for}~~ x< 0 ~\text{and} ~ x>c ~,
\end{cases} 
\end{eqnarray} in order to ensure that the particle in the initial state, above the barrier, has non-zero momentum in accordance with the discussion of the previous section.

 To render the initial state normalized to unity we set the negligibly small overlapping terms

\begin{eqnarray}
\int\mathrm{d}y\, \psi^*_0(y) \psi_j(y) ~, ~~ j = 1, 2, 3, \ldots ~, \nonumber 
\end{eqnarray} equal to zeros. Using this initial state, one obtains for the transition amplitude \eqref{amplitude}

\begin{widetext}
	\begin{eqnarray}\label{amplituda}
	&&\langle\psi_\beta(t=0)|\psi_\beta(t)\rangle =\langle\psi_\beta(t=0)|\widehat{G}_R(t)|\psi_\beta(t=0)\rangle= \nonumber \\&& \frac{1}{\mathcal{Z}}  \left\{ \mathrm{e}^{-\beta \mathcal{E}_0}\int\mathrm{d}x\, \psi^*_0(x)\mathrm{e}^{-it\widehat{H}-\epsilon t}\psi_0(x)  +  \sum_{j=1}^\infty \mathrm{e}^{-\mathcal{E}_j\beta} \int\mathrm{d}x\, \psi^*_j(x)\mathrm{e}^{-it\widehat{H}-\epsilon t}\psi_j(x) \right\} ~, ~~~~~~
	\end{eqnarray}\end{widetext} where $\widehat{G}_R(t)$ denotes the retarded Green's function

\begin{eqnarray}
\widehat{G}_R(t) = \frac{i}{2\pi}\int_{-\infty}^\infty\mathrm{d}\mathcal{E}\, \frac{\mathrm{e}^{-it\mathcal{E}}}{\mathcal{E}-\widehat{H} +i\epsilon} ~, \nonumber 
\end{eqnarray} whence the regularization parameter $\epsilon$ in Eq.\eqref{amplituda}. For our consideration this regularization is not needed, but one should remember that if necessary, this sort of regularization can be safely used. The non-escape probability, Eq.\eqref{krilovfoki}, which is the main tool for our calculations, takes the form
\begin{widetext}
	\begin{eqnarray}\label{neprob}
	\big|\langle\psi_\beta(t=0)|\psi_\beta(t)\rangle\big|^2 = \frac{\mathrm{e}^{-2\beta \mathcal{E}_0}}{\mathcal{Z}^2} \big|\langle\psi_0|\exp\left(-it\widehat{H}\right)|\psi_0\rangle\big|^2 + \frac{1}{\mathcal{Z}^2} \left|\sum_{j=1}^\infty \mathrm{e}^{-\mathcal{E}_j\beta}  \langle\psi_j|\mathrm{e}^{-it\widehat{H}}|\psi_j\rangle\right|^2 + ~~~~~~~~~~~~~~~~~~~~ \nonumber \\ \frac{\mathrm{e}^{-\beta \mathcal{E}_0}}{\mathcal{Z}^2} \langle\psi_0|\exp\left(-it\widehat{H}\right)|\psi_0\rangle\sum_{j=1}^\infty \mathrm{e}^{-\mathcal{E}_j\beta}  \langle\psi_j|\mathrm{e}^{-it\widehat{H}}|\psi_j\rangle^*  + \frac{\mathrm{e}^{-\beta \mathcal{E}_0}}{\mathcal{Z}^2} \langle\psi_0|\exp\left(-it\widehat{H}\right)|\psi_0\rangle^*\sum_{j=1}^\infty \mathrm{e}^{-\mathcal{E}_j\beta}  \langle\psi_j|\mathrm{e}^{-it\widehat{H}}|\psi_j\rangle ~. ~~~~~
	\end{eqnarray}\end{widetext} The first line of Eq.\eqref{neprob} is the sum of tunneling and barrier hopping probabilities and the next line represents the interference between the two phenomena. The piece of physics related to the interference between the (tunnelings in presence of several metastable states and) tunneling and barrier hopping is missing in standard analysis. Let us assume for simplicity that the decay dynamics through the tunneling follows a pure exponential law

\begin{eqnarray}\label{amperti}
\langle\psi_0|\exp\left(-it\widehat{H}\right)|\psi_0\rangle \approx \mathrm{e}^{-it\mathcal{E}_0 - \Gamma t/2} ~. 
\end{eqnarray} For evaluating the statistical sum, which stands for the barrier hopping, let us use the spectrum \eqref{barierzeda}, which allows one to simply calculate the amplitude

\begin{eqnarray}\label{ampori}
	\langle\psi_j|\mathrm{e}^{-it\widehat{H}}|\psi_j\rangle = \mathrm{e}^{-it\mathcal{E}_jt}\begin{cases} \left(1-\frac{k_j t}{mc}\right) ~, ~~  0< t< \frac{mc}{k_j}~,  \\ ~~~~~~~0 ~, ~~~~~~~ t> mc/k_j ~. 
	\end{cases} 
\end{eqnarray} Without alerting the conclusions, one can write the Eq.\eqref{ampori} in analogy to Eq.\eqref{amperti} as

\begin{eqnarray}\label{amp2}
\langle\psi_j|\mathrm{e}^{-it\widehat{H}}|\psi_j\rangle \approx \mathrm{e}^{-it\mathcal{E}_jt - \Gamma_jt/2}~, 
\end{eqnarray} where $\Gamma_j\approx 2k_j/mc$. To simplify matters further, let us employ a low temperature approximation, $\beta\Phi_0 \gg 1$, implying that the statistical sum is dominated by the lower energy terms. In particular, we shall consider just two energy levels: $\mathcal{E}_0$ and $\mathcal{E}_1=\Phi_0$. Substituting now Eqs.(\ref{amperti}, \ref{amp2}) in Eq.\eqref{neprob} and using Eq.\eqref{Lebensdauer} for finding the mean-lifetime, one obtains

\begin{eqnarray}\label{jamuri}
	\tau = \frac{1}{\mathcal{Z}^2}\left( \frac{\mathrm{e}^{-2\beta\mathcal{E}_0}}{\Gamma} + \frac{\mathrm{e}^{-2\beta\Phi_0}}{\Gamma_1} + \right. ~~~~~~~~~~~~~~ \nonumber \\ \left.  \frac{4\mathrm{e}^{-\beta\mathcal{E}_0 -\beta\Phi_0}(\Gamma+\Gamma_1)}{(\Gamma+\Gamma_1)^2+(\Phi_0-\mathcal{E}_0)^2} \right) = \frac{1}{\left(1+\mathrm{e}^{-\beta(\Phi_0-\mathcal{E}_0)}\right)^2}\times \nonumber \\ \left( \frac{1}{\Gamma} + \frac{\mathrm{e}^{-2\beta(\Phi_0-\mathcal{E}_0)}}{\Gamma_1} +   \frac{4\mathrm{e}^{ -\beta(\Phi_0-\mathcal{E}_0)}(\Gamma+\Gamma_1)}{(\Gamma+\Gamma_1)^2+(\Phi_0-\mathcal{E}_0)^2} \right)  ~. ~~~
\end{eqnarray} To put this result in a form that can immediately be applicable to the field theory, let us recall from the previous section that $\Gamma/W(\mathcal{E}_0) \simeq \Gamma_1$, where $W(\mathcal{E}_0) (\ll 1)$ is usually estimated by the bounce solution \cite{Coleman:1977py} 

\begin{eqnarray}
	W(\mathcal{E}_0) = \mathrm{e}^{-\mathfrak{B}(\mathcal{E}_0)} ~. \nonumber 
\end{eqnarray} In the field theory, one usually considers the tunneling from the bottom of a potential well. Thus, we can assume that $\mathcal{E}_0\ll \Phi_0$. For this case, the Eq.\eqref{jamuri} can be put in the form

\begin{eqnarray}\label{oritsevri}
\tau = \frac{1}{\mathcal{Z}^2\Gamma_1}\left( \mathrm{e}^{\mathfrak{B}(\mathcal{E}_0)} + \mathrm{e}^{-2\beta\Phi_0} +   \frac{4\mathrm{e}^{ -\beta\Phi_0}}{1+\Phi^2_0/\Gamma^2_1} \right) ~.
\end{eqnarray} The expression in the brackets is clearly dominated by the first term: $\mathrm{e}^{\mathfrak{B}(\mathcal{E}_0)}$. Because of non-zero temperature, the zero-temperature lifetime, $\mathrm{e}^{\mathfrak{B}(\mathcal{E}_0)}/\Gamma_1$, is now suppressed by the factor $\mathcal{Z}^2$, where

\begin{eqnarray}\label{pirveliori}
	\mathcal{Z}\approx \mathrm{e}^{-\beta\mathcal{E}_0} + \mathrm{e}^{-\beta\Phi_0} ~. 
\end{eqnarray} The maximum value of Eq.\eqref{pirveliori}, which occurs when $\beta \to 0$, is equal to $2$. Thus, at best the Eq.\eqref{oritsevri} can lead to the suppression of the zero-temperature lifetime by the factor $1/4$. However, the approximation \eqref{oritsevri} is valid for small temperatures and the best we can do is to speculate that $\beta \simeq \Phi_0^{-1}$. In this case $\mathcal{Z}\approx 1+\mathrm{e}^{-1} \approx 1.37$ and the suppression is even smaller.

At this point, the reader may wonder how raising the temperature can possibly reduce the lifetime significantly. From the above discussion it is clear that the considerable suppression of the zero-temperature result due to barrier hopping occurs at a temperature so high that $\mathcal{Z}$ includes a large number of over-barrier modes. It is useful to keep in mind that there is an upper bound 

\begin{eqnarray}
	\mathcal{Z} \approx \sum_{j=0}^n\mathrm{e}^{-\beta\mathcal{E}_j} < n+1 ~, \nonumber 
\end{eqnarray} which may be used for a crude estimate of the suppression factor of a zero-temperature lifetime. The number of modes $n$, estimated via $\mathcal{E}_j\lesssim \beta^{-1}$, is simply given by the ratio $c/\beta$ implying that $\mathcal{Z}^2 \simeq c^2/\beta^2$.

Let us now see if the above conceptual framework can lead to any tangible results in field theory.

\section{Field theory applications}

The mean-lifetime derived in the previous section involves a thermal averaging over the amplitudes, Eq.\eqref{neprob}, and the final result represents the thermal average of zero-temperature lifetimes + interference terms, Eqs.(\ref{jamuri}, \ref{oritsevri}). It differs from the existing formalism, which approaches the problem by Boltzmann averaging over the decay rates, see section \ref{semiclassical}. In our discussion, the Eq.\eqref{oritsevri} stands squarely at the crossroads linking the quantum mechanical expression with the field theory one. At this point one may have an objection regarding the Boltzmann averaging over the amplitudes instead of the physical quantity which is of immediate interest for us. We do not want to rebut this objection. Moreover, we also omit the interference terms and just focus on the point which seems more essential and obvious at the same time. The point is that, the correct result for the average lifetime 

\begin{eqnarray}
\tau_\beta = \frac{1}{\mathcal{Z}}\sum_j\mathrm{e}^{-\beta\mathcal{E}_j} \tau(\mathcal{E}_j) \equiv  \frac{1}{\mathcal{Z}}\sum_j\frac{\mathrm{e}^{-\beta\mathcal{E}_j}}{\Gamma(\mathcal{E}_j)}~, \nonumber
\end{eqnarray} cannot be obtained by first averaging the decay rate

\begin{eqnarray}
\Gamma_\beta = \frac{1}{\mathcal{Z}}\sum_j\mathrm{e}^{-\beta\mathcal{E}_j} \Gamma(\mathcal{E}_j) ~, \nonumber 
\end{eqnarray} and then inverting it. 

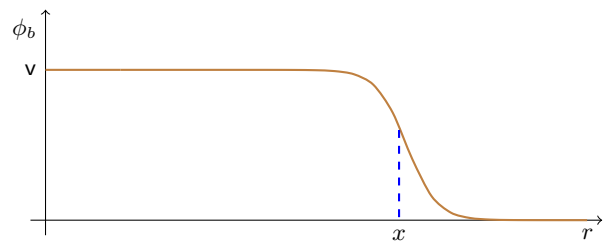
\begin{figure}[h]
	\centering

\begin{tikzpicture}
\draw[->] (-5,0) -- (2.6,0)  node[anchor=north east] {$r$};
\draw[->] (-4.8,-0.2) -- (-4.8,2.8)  node[anchor=north east] {$\phi_b$};
\draw[scale=1,domain=-3.8:2.4,smooth,variable=\x,brown, thick] plot ({\x}, 
{2-1*(1+tanh(2.5*\x))});
\draw[scale=1,domain=-4.8:-3.8,smooth,variable=\x,brown, thick] plot ({\x}, 
{2});

\node at (-5,2) {$\mathsf{v}$};
\draw [blue, thick, dashed] (-0.1,1.2) -- (-0.1,0) node[black, below]{$x$} ;

\end{tikzpicture}

	\caption{The bubble profile.} \label{fig2}
\end{figure}

To apply the above arguments to the field theory, which may be defined by the Lagrangian density

\begin{eqnarray}\label{velislag}
\mathcal{L} = \frac{\partial_\alpha\phi\partial^\alpha\phi}{2} - \frac{\lambda\phi^2}{4}\Big(\phi - \mathsf{v}\Big)^2 +  \frac{3\epsilon\phi}{4\pi \mathsf{v}} ~ ,
\end{eqnarray} we use the formalism of the thin-walled-bubble Ansatz \cite{Kobzarev:1974cp, Voloshin:1993ks, Coleman:1977py}, see Fig.\ref{fig2}.

\begin{figure}[h]
	\centering
	
	\begin{tikzpicture}
	\draw[->] (-0.2,0) -- (7,0)  node[anchor=north east] {$x$};
	\draw[->] (-0,-0.2) -- (-0,3.2)  node[anchor=north east] {$\Phi(x)$};
	\draw[scale=1,domain=0:5.7,smooth,variable=\x,brown, thick] plot ({\x}, 
	{0.55*\x*\x -0.1*\x*\x*\x});
	\draw [blue, thick, dashed] (0,1.2) -- (1.8,1.2)  node[black, pos=0.5, above] {Subcritical};
	\draw [blue, thick, dashed] (5.1,1.2) -- (6.5,1.2) node[black, pos=0.48, above] {Critical}  node[black, right] {$\mathcal{E}$};
	\draw [blue, thick, dashed] (0,2.46) -- (5,2.46)   node[black, right] {$\Phi_0$};
	\draw [blue, thick, dashed] (3.7,2.45) -- (3.7,0) node[black, below]{$x_{top}$} ;
	\draw [blue, thick, dashed] (1.8,1.2) -- (1.8,0) node[black, below]{$c(\mathcal{E})$} ;

	\end{tikzpicture}

	\caption{Potential for the one-dimensional model.} \label{fig3}
\end{figure}
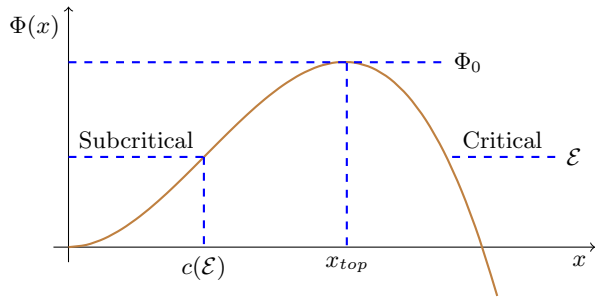

\noindent In view of the Lagrangian \eqref{velislag}, bubble can be approximated by

\begin{eqnarray}
	\phi_b = \frac{\mathsf{v}}{2}\left(1-\tanh\left[\frac{\mathsf{v}\sqrt{\lambda}(r-x)}{2\sqrt{2}}\right] \right) ~. \nonumber 
\end{eqnarray} Thus, the bubble is characterized with the radius $x$, which is a dynamical quantity, and with a thin wall heaving the thickness of the order of $1/v\sqrt{\lambda}$. This Ansatz reduces the field-theory problem to the one-dimensional mechanical one with the Lagrangian\footnote{Strictly speaking, this approximation breaks down when the bubble radius becomes of the order of the wall thickness:  $x \sim 1/\mathsf{v}\sqrt{\lambda}$.} \cite{Bitar:1978vx, 1978PhRvD..17.1056K, Michel:2019nwa}

\begin{eqnarray}\label{redutsirebuli}L = 4\pi\int_0^\infty\mathrm{d}r\, r^2 \frac{\dot{\phi}_b^2}{2}  -
4\pi\int_0^\infty\mathrm{d}r\, r^2 \left(	\frac{\nabla\phi_b\cdot\nabla\phi_b}{2} +U(\phi_b)\right)  \nonumber \\ = 2\dot{x}^2\pi\int_0^\infty\mathrm{d}r\, r^2 \left(\frac{\mathrm{d}\phi_b}{\mathrm{d}x}\right)^2 - \frac{4\pi x^3}{3}U(\mathsf{v}) - ~~~~~~~~~~ \nonumber \\    4\pi x^2 \int_0^\mathsf{v}\mathrm{d}\phi \, \sqrt{2U(\phi)} \equiv  \frac{\mu(x)\dot{x}^2}{2} +\epsilon x^3 - \sigma x^2 ~,~~~~~  
\end{eqnarray} where

\begin{eqnarray}
\mu(x) = \frac{\sqrt{2}\pi \mathsf{v}^3\sqrt{\lambda}x^2}{3} ~~ \text{and}~~ \sigma = \sqrt{\frac{\lambda}{2}} \frac{2\pi \mathsf{v}^3}{3}  ~.\nonumber 
\end{eqnarray} The potential for this one-dimensional problem has the form shown in Fig.\ref{fig3}.

\noindent The analogy with the one-dimensional problem helps to clarify the picture. A bubble of true vacuum of the subcritical size nucleated in a surrounding sea of the false vacuum will just oscillate before it will tunnel \cite{Voronov:1976ae, Bogolyubsky:1976nx, Belova:1976ui, Copeland:1995fq}. On the other hand, a bubble of the critical size will spread forever - converting false vacuum to true \cite{Coleman:1977py, Kobzarev:1974cp}. The bounce solution \cite{Coleman:1977py}, which corresponds to the $\mathcal{E} = 0$ case in Fig.\ref{fig3}, determines the probability - denoted by $W(\mathcal{E})$ in section \ref{semiclassical}, rather than a decay rate. In order to infer a rate, we need to identify a characteristic time as it was done in section \ref{semiclassical}. For the subcritical bubbles this time scale is naturally defined by the Eq.\eqref{sikhshire}. Denoting by $\rho(\mathcal{E})\mathrm{d}\mathcal{E}$ the nucleation probability of the subcritical bubble in the false vacuum in the energy range $(\mathcal{E}, \mathcal{E}+\mathrm{d}\mathcal{E})$, one can express the lifetime of false vacuum as

\begin{eqnarray}
	\tau = \int_0^{\Phi_0}\frac{\mathrm{d}\mathcal{E}\rho(\mathcal{E})}{W(\mathcal{E})}  \int_0^{c(\mathcal{E})}\mathrm{d}x\, \sqrt{\frac{\mu(x)}{ 2\big(\mathcal{E}-\Phi(x)\big)}} ~. \nonumber 
\end{eqnarray} Thus, in analogy with section \ref{semiclassical}, the subcritical bubbles play the role of the metastable states. Unfortunately, we do not have an elegant systematic techniques for estimating the nucleation probability of the subcritical bubble \cite{Voloshin:1993ks}. What we can say on general grounds is that $\rho(\mathcal{E})$ decreases rapidly with $\mathcal{E}$ leading to the idea that the bounce may provide a fairly good approximation. At the finite temperature, however, $\rho(\mathcal{E})$ is simply given by the Boltzmann factor $\mathrm{e}^{-\beta\mathcal{E}}/\mathcal{Z}$. Correspondingly, the lifetime of the false vacuum which decays through the tunneling at a finite temperature is given by

\begin{eqnarray}\label{thermaltunneling}
\tau = \frac{1}{\mathcal{Z}} \int_0^{\Phi_0}\frac{\mathrm{d}\mathcal{E}\mathrm{e}^{-\beta\mathcal{E}}}{W(\mathcal{E})}  \int_0^{c(\mathcal{E})}\mathrm{d}x\, \sqrt{\frac{\mu(x)}{ 2\big(\mathcal{E}-\Phi(x)\big)}} ~, 
\end{eqnarray} where

\begin{eqnarray}\label{mamravli}
\mathcal{Z} = \int_0^{\Phi_0}\mathrm{d}\mathcal{E}\, \mathrm{e}^{-\beta\mathcal{E}} = \frac{1-\mathrm{e}^{-\beta\Phi_0}}{\beta} ~. 
\end{eqnarray}

 The bubble solution with the energy $\Phi_0$ represents the saddle point of the potential 

\begin{eqnarray}
V[\phi] = \int\mathrm{d}^3x \, \left(\frac{\nabla\phi\cdot\nabla\phi}{2} +U(\phi)\right) ~, \nonumber 
\end{eqnarray} and thus determines the height of the potential barrier \cite{Affleck:1980ac, Linde:2005ht}. Correspondingly, the probability for barrier hopping is given by $\mathrm{e}^{-\beta\Phi_0}/\mathcal{Z}$. From Eq.\eqref{redutsirebuli} one finds that 

\begin{eqnarray}\label{simaghle}
	\Phi_0 = \frac{\sigma}{3}\left(\frac{2\sigma}{3\epsilon}\right)^2 ~. 
\end{eqnarray} In the high ($\beta\Phi_0\ll 1$) and low ($\beta\Phi_0\gg 1$) temperature limits, the Eq.\eqref{mamravli} gives $\mathcal{Z}\approx \Phi_0$ and $\mathcal{Z}\approx 1/\beta$, respectively. As to the Eq.\eqref{thermaltunneling}, one can estimate it in the high and low temperature asymptotic regimes as follows. The expression

\begin{eqnarray}
	\frac{\mathrm{e}^{-\beta\mathcal{E}}}{W(\mathcal{E})} \equiv \mathrm{e}^{-\beta\mathcal{E}+\mathfrak{B}(\mathcal{E})} ~, \nonumber 
\end{eqnarray} is monotonically decreasing with energy. On the other hand, the expression

\begin{eqnarray}\label{droisintegrali}
\int_0^{c(\mathcal{E})}\mathrm{d}x\, \sqrt{\frac{\mu(x)}{ 2\big(\mathcal{E}-\Phi(x)\big)}} ~, 
\end{eqnarray} is increasing with energy but it varies much slowly relative to the exponential factor. The integral \eqref{droisintegrali} diverges for $\mathcal{E}=\Phi_0$ and is zero for $\mathcal{E}=0$. To avoid this divergence and, on the other hand, to set the time scale for the decay in a low-temperature limit, one can use the cutoff that naturally exists in the model. Namely, the variable $x$ is defined up to the wall thickness and one can replace the upper limit of the integral by $x_{\wedge}\equiv x_{top}-1/\mathsf{v}\sqrt{\lambda}$ and the lower limit by $x_{\vee}\equiv 1/\mathsf{v}\sqrt{\lambda}$. Stated more precisely, one can identify the corresponding energy scales 

\begin{eqnarray}
	\mathcal{E}_{\vee, \wedge} = \sigma x^2_{\vee, \wedge} - \epsilon x^3_{\vee, \wedge} ~, \nonumber 
\end{eqnarray} and modify the Eq.\eqref{thermaltunneling} as follows

\begin{eqnarray}
\tau = \frac{1}{\mathcal{Z}} \int_0^{\mathcal{E}_{\vee}}\frac{\mathrm{d}\mathcal{E}\mathrm{e}^{-\beta\mathcal{E}}}{W(\mathcal{E})}  \int_0^{x_{\vee}}\mathrm{d}x\, \sqrt{\frac{\mu(x)}{ 2\big(\mathcal{E}_{\vee}-\Phi(x)\big)}} + \nonumber \\  \frac{1}{\mathcal{Z}} \int_{\mathcal{E}_{\vee}}^{\mathcal{E}_{ \wedge}}\frac{\mathrm{d}\mathcal{E}\mathrm{e}^{-\beta\mathcal{E}}}{W(\mathcal{E})}  \int_{0}^{c(\mathcal{E})}\mathrm{d}x\, \sqrt{\frac{\mu(x)}{ 2\big(\mathcal{E}-\Phi(x)\big)}} + \nonumber \\ \frac{1}{\mathcal{Z}} \int_{\mathcal{E}_{\wedge}}^{\Phi_0}\frac{\mathrm{d}\mathcal{E}\mathrm{e}^{-\beta\mathcal{E}}}{W(\mathcal{E})}  \int_0^{x_{\wedge}}\mathrm{d}x\, \sqrt{\frac{\mu(x)}{ 2\big(\mathcal{E}_{\wedge}-\Phi(x)\big)}} ~. \nonumber 
\end{eqnarray} In the low-temperature limit one will have 

\begin{eqnarray}
	\tau \simeq \frac{1}{W(\mathcal{E}=0)} \int_0^{x_{\vee}}\mathrm{d}x\, \sqrt{\frac{\mu(x)}{ 2\big(\mathcal{E}_{\vee}-\Phi(x)\big)}} ~. \nonumber
\end{eqnarray} In the high-temperature limit, one can safely omit the factor: $\mathrm{e}^{-\beta \mathcal{E}} \approx 1$. Besides, we know that $1/W(\mathcal{E}) = \exp\big(\mathfrak{B}(\mathcal{E})\big)$ is sharply peaked at $\mathcal{E}=0$ \cite{Coleman:1977th}, however, we do not know its localization width. For this reason, we use the energy scale $\mathcal{E}_\vee$ in the role of the width that results in the approximate result   

\begin{eqnarray}
\tau \simeq \frac{\mathcal{E}_\vee}{\Phi_0} \frac{1}{W(\mathcal{E}=0)} \int_0^{x_{\vee}}\mathrm{d}x\, \sqrt{\frac{\mu(x)}{ 2\big(\mathcal{E}_{\vee}-\Phi(x)\big)}}~, 
\end{eqnarray} which clearly indicates that the only suppression factor (as compared to the zero-temperature case) is $\mathcal{E}_\vee/\Phi_0$. In the thin-wall approximation $\Phi_0$ is large enough, see Eq.\eqref{simaghle}, and correspondingly, this suppression factor becomes quite appreciable.

The discussion so far almost precisely parallels the quantum-mechanical picture of the previous sections. To proceed in the same way, the over-barrier states can be identified either with multi-bubble configurations implying the energy levels $2\Phi_0, 3\Phi_0$ an so forth, or one can propose a quantum-mechanical description of the bubbles (like it was suggested in \cite{Michel:2019nwa}) and consider it apt that due to quantum fluctuations of the wall position, $x$, there should exist bubbles with the same $x$ but with discretely increasing wall energy. That is, in the latter case one would have the bubbles with radius $x_{top}$ and the over-barrier spectrum $\propto 1/x_{top}$ (similar to what was consider in the previous section). In view of the discussion of the previous section, one infers that the existence of such bubbles allows one to estimate the false vacuum lifetime at a finite temperature as\footnote{Here we omit the thermal corrections to the tunneling given by Eq.\eqref{thermaltunneling} as it is expected to be less significant for decreasing the lifetime.}

\begin{eqnarray}\label{dziritadi}
\tau \simeq \frac{\beta}{x_{top}} \, \frac{\mathrm{e}^{\mathfrak{B}(\mathcal{E}=0)}}{\Gamma_1} ~. 
\end{eqnarray} One sees that if $x_{top}$ is large enough, then the lifetime can be reduced significantly even at relatively low temperatures. The value of $x_{top}$ in the thin-wall approximation (see Eq.\eqref{redutsirebuli}) is large enough 

\begin{eqnarray}
	x_{top} = \frac{2\sigma}{3\epsilon} ~~\Rightarrow~~  \tau \simeq \frac{3\epsilon\beta}{2\sigma} \, \frac{\mathrm{e}^{\mathfrak{B}(\mathcal{E}=0)}}{\Gamma_1} ~. \nonumber 
\end{eqnarray} 

In real physical models, where $x_{top}$ is not large, this suppression mechanism is merely useless. For instance, it cannot affect the lifetime of the Higgs vacuum \cite{Sher:1988mj, Anderson:1990aa, Arnold:1991cv} or the time-scale of anomalous electroweak baryon number violating process \cite{Kuzmin:1985mm}.

\section{Summary}

1. Starting point of our discussion is the observation that the lifetime of an unstable system is a random variable and its consistent description requires the knowledge of a distribution function. 

2. Further observation is that the Krylov-Fock non-escape probability provides such a distribution function. However, besides the mean-lifetime that naturally follows from this distribution function, Eq.\eqref{Lebensdauer}, there exists an alternative definition, Eq.\eqref{AlterAusdruck}, and it would be desirable to check experimentally which of them can be considered as a reliable one. For this reason, in section \ref{Cauchy-Lorentz} we have worked out both of these quantities for the Cauchy-Lorentz energy distribution function. 

3. And here comes the next, unpleasant, observation that for the exponential decay both definitions of the mean-lifetime imply remarkably large fluctuations. As manifested by the standard discussion of the radioactive decay, see section \ref{fluctuations}, this problem maybe somewhat more generic for the unstable systems. An obvious downside of this fact is that it considerably reduces the predictive power of the false vacuum lifetime. In general, it should be emphasized that an obvious deficiency of the existing quantum-mechanical as well as field-theory estimates of the mean-lifetime of an unstable system is the lack of its fluctuation rate (or uncertainty). It should not be confused with the uncertainties that might be related to the imprecisions of the input parameters of the model.        

4. As a logical continuation of our discussion, we address in the next section a finite temperature decay in terms of the Krylov-Fock non-escape probability. For this purpose we have used thermo-field dynamics formalism. The new features as compared to the existing descriptions are as follows. The amplitudes of tunneling corresponding to different metastable states interfere with each other and also with the amplitude of the barrier hopping. It is just a manifestation of the double-slit phenomenon in quantum mechanics. If the particle can escape from the potential well through the different "slits" (as is the case at the finite temperature), then one should naturally expect the interference between the amplitudes. Correspondingly, we shall have an additional contribution to the mean-lifetime.          

And the other feature is that the thermal averaging (with respect to the Boltzmann distribution) is over the lifetimes. Of course, it would be incorrect to find first the average value of the decay rate

\begin{eqnarray}
	\Gamma_\beta = \frac{1}{\mathcal{Z}} \sum \mathrm{e}^{-\beta\mathcal{E}_j}\Gamma(\mathcal{E}_j) ~, \nonumber
\end{eqnarray} and then estimate the mean-lifetime as $\tau_\beta = 1/ \Gamma_\beta $. If we are interested in the lifetime at a finite temperature, the natural approach would be, of course, to write 

\begin{eqnarray}
	\tau_\beta = \frac{1}{\mathcal{Z}} \sum \mathrm{e}^{-\beta\mathcal{E}_j}\tau(\mathcal{E}_j) ~. \nonumber
\end{eqnarray} In the simplest case, $\tau(\mathcal{E}_j) = 1/\Gamma(\mathcal{E}_j)$, the difference between these two expressions is obvious. However, there
is a subtle point noticed by Referee. Namely, it would be desirable for the mean-lifetime to have natural low- and high-temperature limits. That is, if in the low-temperature limit the tunneling contributes essentially to the mean-lifetime, in the high-temperature limit one
would naturally expect the mean-lifetime to be almost independent of it as over-barrier jumping becomes the dominant process. The mean-lifetime defined as

\begin{eqnarray}
	\tau_\beta = \frac{1}{\Gamma_\beta} = \frac{1}{\Gamma_{tun}+\Gamma_{ob} } ~, \nonumber 
\end{eqnarray} where we have used notations similar to Eqs.(\ref{ftmpep}, \ref{ftmph}), shows the desired asymptotic behavior - in the limit $\beta\to\infty$ it is dominated by the tunneling decay rate, while in the high-temperature limit $\Gamma_{tun}$ becomes negligible as compared to $\Gamma_{ob}$ and does not contribute to the mean-lifetime. The result obtained by the thermo-field approach involves the contribution from tunneling even
in the high-temperature limit, see Eq.\eqref{oritsevri}. This peculiarity can be understood better by considering the limiting cases of the Boltzmann average 

\begin{eqnarray}
\tau_\beta  = \frac{1}{\mathcal{Z}} \sum \mathrm{e}^{-\beta \mathcal{E}_j}\tau(\mathcal{E}_j) = \frac{1}{\mathcal{Z}}\sum \frac{ \mathrm{e}^{-\beta \mathcal{E}_j}}{\Gamma(\mathcal{E}_j)} ~. \nonumber 
\end{eqnarray} One sees that the low-temperature limit $\beta \to \infty$ singles out the metastable state $\mathcal{E}_0$, but in the limit $\beta\to 0 $ the sum takes the form

\begin{eqnarray}
\frac{1}{\mathcal{Z}} \left\{\frac{1}{\Gamma(\mathcal{E}_0)} + \frac{1}{\Gamma(\mathcal{E}_1)} +\cdots \right\} ~, \nonumber 
\end{eqnarray} where $\Gamma(\mathcal{E}_0)$ is the tunneling probability and $\Gamma(\mathcal{E}_1)$ stands for the barrier hopping one. $\Gamma(\mathcal{E}_1)$ is temperature dependent and increases with temperature. That is how $1/\Gamma(\mathcal{E}_0)$ survives in the high temperature limit. It would be really interesting to check experimentally as closely as possible the predictions of the thermo-field dynamics for
the decay of an unstable systems. It seems quite feasible to formulate such experiments.

5. The final section is devoted to the field theory applications of the above results. Thin wall approximation reduces the problem to the one-dimensional case that allows one to carry out the discussion more or less straightforwardly. Certainly, in contrast to the quantum mechanics, in field theory our knowledge about the metastable states is somewhat restricted. One usually thinks in terms of the probabilities estimated by using $O(4)$ and $O(3)$ symmetric bounces. The former one determines the field tunneling probability from the bottom of the potential and the latter one allows one to determine the height of the potential barrier that is important for estimating the barrier hopping probability at a finite temperature. Judging in terms of these solutions, one can infer certain conclusions regarding the lifetime of a false vacuum.

\begin{acknowledgments}
We would like to thank Zurab Kepuladze and George Lavrelashvili for helpful discussions. This research was supported in part by the Rustaveli National Science Foundation of Georgia under Grant No. FR-19-8306.
\end{acknowledgments}

\appendix*

\section{Evaluating integrals}

For evaluating the integral \eqref{amplitude} for the Cauchy-Lorentz distribution, it is convenient to introduce a dimensionless variable $\xi \equiv 2(\mathcal{E}-\mathcal{E}_0)/\Gamma$   

\begin{widetext}
	\begin{eqnarray}\label{asympgamma}
	\int_0^\infty \frac{\mathrm{d}\mathcal{E}\, \Gamma\exp(-i\mathcal{E}t)}{\left(\mathcal{E} - \mathcal{E}_0\right)^2 + \Gamma^2/4} =\frac{2\exp(-i\mathcal{E}_0t)}{\Gamma} \int_{-\xi_0}^\infty \frac{\mathrm{d}\xi  \, \exp(-it\xi\Gamma )}{\xi^2+1} =  \frac{2\exp(-i\mathcal{E}_0t)}{\Gamma} \int_{-\xi_0}^\infty \frac{\mathrm{d}\xi  \, \exp(-it\xi\Gamma /2)}{\xi^2+1}  =  \nonumber \\  \frac{2\exp(-i\mathcal{E}_0t)}{\Gamma} \int_{-\infty}^\infty \frac{\mathrm{d}\xi  \, \exp(-it\xi\Gamma /2)}{\xi^2+1} - \frac{2\exp(-i\mathcal{E}_0t)}{\Gamma} \int_{-\infty}^{-\xi_0} \frac{\mathrm{d}\xi  \, \exp(-it\xi\Gamma /2)}{\xi^2+1}  = \nonumber \\   \frac{\exp(-i\mathcal{E}_0t - \Gamma t /2)}{\Gamma} -  \frac{2\exp(-i\mathcal{E}_0t)}{\Gamma} \int_{-\infty}^{-\xi_0} \frac{\mathrm{d}\xi  \, \exp(-it\xi\Gamma /2)}{\xi^2+1} ~. ~~~~~~~~
	\end{eqnarray} Here $\xi_0 \equiv 2\mathcal{E}_0/\Gamma$. From Eq.\eqref{asympgamma} one sees that for $\xi_0\gg 1$, the second term determining deviation from the exponential decay becomes smaller. For estimating the order of magnitude of the deviation, one can expand this term in powers of $1/\xi_0$ by using repeated integration by parts

\begin{eqnarray}
\int_{-\infty}^{-\xi_0} \frac{\mathrm{d}\xi  \, \exp(-it\xi\Gamma /2)}{\xi^2+1} = \frac{2i\exp(it\xi_0\Gamma/2)}{t\Gamma(1+\xi_0^2)} +  \frac{8\xi_0\exp(it\xi_0\Gamma/2)}{t^2\Gamma^2(1+\xi_0^2)^2} + \frac{8}{t^2\Gamma^2}\int_{-\infty}^{-\xi_0}\mathrm{d}\xi \, \frac{(1-3\xi^2)\exp(-it\xi\Gamma /2)}{(1+\xi^2)^3} ~, \nonumber 
\end{eqnarray} where the remainder term is bounded as

\begin{eqnarray}
\left|\int_{-\infty}^{-\xi_0}\mathrm{d}\xi \, \frac{(1-3\xi^2)\exp(-it\xi\Gamma /2)}{(1+\xi^2)^3}\right| \leq   \int_{-\infty}^{-\xi_0}\mathrm{d}\xi \, \frac{|1-3\xi^2|}{(1+\xi^2)^3}  = \frac{\xi_0}{(1+\xi_0^2)^2} ~. \nonumber
\end{eqnarray} One can, of course, continue the expansion to any order in $1/\xi_0$. For our discussion, it is important to estimate the asymptotic behavior of Eq.\eqref{asympgamma} as $t\to\infty$. For this purpose, let us introduce a new variable $\eta = \xi t$ and use again repeated integration by parts

\begin{eqnarray}\label{asimptotika}
\int_{-\xi_0}^\infty \frac{\mathrm{d}\xi  \, \exp(-it\xi\Gamma /2)}{\xi^2+1} = t \int_{-t\xi_0}^{\infty}\mathrm{d}\eta \, \frac{\exp(-i\eta \Gamma/2)}{t^2+\eta^2} =  -\frac{2i\exp(it\xi_0\Gamma/2)}{t\Gamma (1+\xi_0^2)} - \nonumber \\ \frac{8\xi_0 \exp(it\xi_0\Gamma/2)}{\Gamma^2 t^2(1+\xi_0^2)^2} + \frac{8}{\Gamma^2t^2}   \int_{-\xi_0}^\infty\mathrm{d}\xi \, \frac{(1-3\xi^2)\exp(-it\xi\Gamma/2)}{(1+\xi^2)^3} ~. 
\end{eqnarray} The order of magnitude of the last integral in Eq.\eqref{asimptotika} can be easily estimated (it is assumed that $\xi_0>1/\sqrt{3}$)

\begin{eqnarray}
\left|\int_{-\xi_0}^\infty\mathrm{d}\xi \, \frac{(1-3\xi^2)\exp(-it\xi\Gamma/2)}{(1+\xi^2)^3}\right| \leq \sqrt{3} - \frac{\xi_0}{(1+\xi_0^2)^2} ~. \nonumber
\end{eqnarray} One sees that for large values of $t$, the integral \eqref{asympgamma} decays at least as $t^{-1}$ and, correspondingly, the quantity $\omega(t)$ will decay at least as $t^{-2}$. Therefore, in Eq.\eqref{Lebensdauer} one can safely ignore the term $t\omega(t)$ in the limit $t\to\infty$. Therefore, for the mean lifetime one obtains

	\begin{eqnarray}\label{sashualo}
	\tau =  \frac{N^2}{4\pi^2\Gamma^2} \int_0^\infty\mathrm{d}t \left(\exp(- \Gamma t /2) - 2\int_{-\infty}^{-\xi_0} \frac{\mathrm{d}\xi  \, \cos(t\xi\Gamma /2)}{\xi^2+1}\right)^2 + \frac{N^2}{4\pi^2\Gamma^2} \int_0^\infty\mathrm{d}t \left(2\int_{-\infty}^{-\xi_0} \frac{\mathrm{d}\xi  \, \sin(t\xi\Gamma /2)}{\xi^2+1}\right)^2 =   \frac{N^2}{4\pi^2\Gamma^3} - \nonumber \\ \frac{2N^2}{\pi^2\Gamma^3}\int_{-\infty}^{-\xi_0} \frac{\mathrm{d}\xi}{(1+\xi^2)^2}+ \frac{N^2}{\pi^2\Gamma^2}\int_{-\infty}^{-\xi_0} \frac{\mathrm{d}\xi_1}{\xi_1^2+1}\int_{-\infty}^{-\xi_0} \frac{\mathrm{d}\xi_2}{\xi_2^2+1}  \, \int_0^\infty\mathrm{d}t\, \cos\left(\frac{t\Gamma}{2}(\xi_1-\xi_2)\right) = \frac{N^2}{4\pi^2\Gamma^3} -  \frac{2N^2}{\pi^2\Gamma^3}\int_{-\infty}^{-\xi_0} \frac{\mathrm{d}\xi}{(1+\xi^2)^2} + \nonumber \\ \frac{N^2}{2\pi^2\Gamma^2}\int_{-\infty}^{-\xi_0} \frac{\mathrm{d}\xi_1}{\xi_1^2+1}\int_{-\infty}^{-\xi_0} \frac{\mathrm{d}\xi_2}{\xi_2^2+1}  \, \int_{-\infty}^\infty\mathrm{d}t\, \exp\left(\frac{it\Gamma}{2}(\xi_1-\xi_2)\right) =  \frac{N^2}{4\pi^2\Gamma^3}  +  \frac{2N^2(\pi-1)}{\pi^2\Gamma^3}\int_{-\infty}^{-\xi_0} \frac{\mathrm{d}\xi}{(1+\xi^2)^2} = \nonumber \\ \frac{N^2}{4\pi^2\Gamma^3} + \frac{N^2(\pi-1)}{\pi^2\Gamma^3}\left(\arctan(-\xi_0) - \frac{\xi_0}{1+\xi_0^2}+\frac{\pi}{2}\right) = \frac{N^2}{4\pi^2\Gamma^3} + \frac{N^2(\pi-1)}{\pi^2\Gamma^3}\left(\frac{2}{3\xi_0^3} - \frac{4}{5\xi_0^5}+\frac{6}{7\xi_0^7} +\cdots\right)  ~.
	\end{eqnarray} Let us now evaluate $\widetilde{\tau}$. For this we need $\omega(t)$ that can be read off without much trouble from Eq.\eqref{sashualo}. Hence, we find 
	
	\begin{eqnarray}\label{integrali}
	\widetilde{\tau}\tau =	\int_0^\infty\mathrm{d}t \, t\omega(t) = \frac{N^2}{4\pi^2\Gamma^2} \int_0^\infty\mathrm{d}t \, t\exp(-\Gamma t) - \frac{N^2}{\pi^2\Gamma^2}\int_{-\infty}^{-\xi_0}  \frac{\mathrm{d}\xi}{1+\xi^2} \int_0^\infty\mathrm{d}t \, t\exp\left(-\frac{\Gamma t}{2}\right)\cos\left(\frac{t\xi\Gamma}{2}\right) + \frac{N^2}{\pi^2\Gamma^2}\int_{-\infty}^{-\xi_0}  \frac{\mathrm{d}\xi_1}{1+\xi_1^2} \times \nonumber \\ \int_{-\infty}^{-\xi_0}  \frac{\mathrm{d}\xi_2}{1+\xi_2^2} \int_0^\infty\mathrm{d}t \, t\cos\left(\frac{t\Gamma(\xi_1-\xi_2)}{2}\right) = \frac{N^2}{4\pi^2\Gamma^4} - \frac{4N^2}{\pi^2\Gamma^4}\int_{-\infty}^{-\xi_0}  \frac{\mathrm{d}\xi (1-\xi^2)}{(1+\xi^2)^3}  -  \frac{4N^2}{\pi^2\Gamma^4}\int_{-\infty}^{-\xi_0}  \frac{\mathrm{d}\xi_1}{1+\xi_1^2}\int_{-\infty}^{-\xi_0}  \frac{\mathrm{d}\xi_2}{1+\xi_2^2} \frac{1}{(\xi_1-\xi_2)^2}  ~.~~~~ 
	\end{eqnarray}\end{widetext} Here we have used 

\begin{eqnarray}
\int_0^\infty\mathrm{d}t \, t\cos(\alpha t) = \frac{\mathrm{d}}{\mathrm{d}\alpha}  \int_0^\infty\mathrm{d}t \, \sin(\alpha t) = \nonumber \\ \frac{\mathrm{d}}{\mathrm{d}\alpha} \Im \int_{-\infty}^\infty\mathrm{d}t \,\theta(t) \mathrm{e}^{i\alpha t} = -\frac{1}{\alpha^2}  ~, \nonumber  
\end{eqnarray} which follows from the well known integral representation of the step-function     

\begin{eqnarray}
\theta(t) = \frac{i}{2\pi}\int_{-\infty}^\infty\mathrm{d}\alpha \, \frac{\mathrm{e}^{-i\alpha t}}{\alpha +i\epsilon} ~. \nonumber
\end{eqnarray} It is equivalent to using the following redefinition for making the integral convergent 

\begin{eqnarray}
\int_0^\infty\mathrm{d}t \, t \cos(\alpha t) ~\to ~   \int_0^\infty\mathrm{d}t \, t \cos(\alpha t) \mathrm{e}^{-\epsilon t} = \nonumber \\ -\Re\frac{\mathrm{d}}{\mathrm{d}\epsilon}\int_0^\infty\mathrm{d}t \, \mathrm{e}^{i\alpha t -\epsilon t} = \frac{1}{(\epsilon - i\alpha)^2} ~. ~~~~~~~ \nonumber
\end{eqnarray} In the last integral in Eq.\eqref{integrali}, however, we omitted $\epsilon$ term - tacitly assuming that because of singularity occurring at $\xi_1=\xi_2$, this integral has to be interpreted by a suitable subtraction of divergences. For this purpose we shall use somewhat different regularization. First, with the use of Wolfram Mathematica, let us carry out the integration with respect to $\xi_2$ from the regularized expression
\begin{widetext}
	\begin{eqnarray}
	\int^{\infty}_{\xi_0}  \frac{\mathrm{d}\xi_2}{1+\xi_2^2} \frac{1}{(\xi_1-\xi_2)^2+\epsilon^2} =  \frac{\epsilon^2 \arctan\left(\frac{\xi_1-\xi_2}{\epsilon}\right)+\epsilon \left(\epsilon^2+\xi_1^2-1\right) \arctan(\xi_2)}{\epsilon \left(\epsilon^4+2 \epsilon^2 \left(\xi_1^2-1\right)+\left(1+\xi_1^2\right)^2\right)} +\nonumber \\  \left.	\frac{\arctan\left(\frac{\xi_2-\xi_1}{\epsilon}\right)+\xi_1^2
		\arctan\left(\frac{\xi_2-\xi_1}{\epsilon}\right)+\epsilon \xi_1 \ln\left(\frac{1+\xi_2^2}{\epsilon^2+(\xi_1-\xi_2)^2}\right)}{\epsilon \left(\epsilon^4+2 \epsilon^2 \left(\xi_1^2-1\right)+\left(1+\xi_1^2\right)^2\right)}\right|^{\xi_2=\infty}_{\xi_2=\xi_0} ~. 
	\end{eqnarray}\end{widetext} From this expression we drop the terms that are either divergent or vanishing in the limit $\epsilon\to 0$. Accordingly, we shall have   

\begin{eqnarray}
\int^{\infty}_{\xi_0}  \frac{\mathrm{d}\xi_2}{1+\xi_2^2} \frac{1}{(\xi_1-\xi_2)^2+\epsilon^2} = \frac{\pi \left(\xi_1^2-1\right)}{2\left(1+\xi_1^2\right)^2} - \nonumber \\ \frac{\xi_1 \ln\left(\frac{1+\xi_0^2}{\epsilon^2+(\xi_1-\xi_0)^2}\right)}{\left(1+\xi_1^2\right)^2} - \frac{\left(\xi_1^2-1\right) \arctan(\xi_0)}{\left(1+\xi_1^2\right)^2}  ~. ~~~~~~~~\nonumber 
\end{eqnarray} Integrating further with respect to $\xi_1$ and dropping again the terms either diverging or vanishing when $\epsilon\to 0$, one obtains

\begin{eqnarray}
\int^{\infty}_{\xi_0}  \frac{\mathrm{d}\xi_1}{1+\xi_1^2}\int^{\infty}_{\xi_0}  \frac{\mathrm{d}\xi_2}{1+\xi_2^2} \frac{1}{(\xi_1-\xi_2)^2} = \nonumber \\ \left(\frac{\pi}{2} - \arctan(\xi_0)\right)\frac{2\xi_0}{1+\xi_0^2} ~. \nonumber 
\end{eqnarray}

\end{document}